\documentclass[double-column,10pt]{IEEEtran}
\usepackage[utf8]{inputenc}
\usepackage{graphicx}
\usepackage{amsmath}
\usepackage{mathrsfs}
\usepackage{amssymb}
\usepackage{bm}
\usepackage{hyperref}
\usepackage{booktabs}
\usepackage[numbers]{natbib}

\def\be{\begin{equation}}
\def\ee{\end{equation}}
\newcommand{\refig}[1]{Fig.~\ref{#1}}
\newcommand{\reftab}[1]{Table~\ref{#1}}
\newcommand{\tx}[1]{\textrm{#1}}

\newcommand{\modela}{H256r}		
\newcommand{\modelB}{Ml512r}		
\newcommand{\modelC}{Mm512r}		
\newcommand{\modelD}{Mh512r}		
\newcommand{\modelA}{H512m1}		
\newcommand{\modelBm}{Ml512m1}  	

\title{ECHO-3DHPC: relativistic accretion disks onto black holes}
\author{\IEEEauthorblockN{Matteo Bugli\IEEEauthorrefmark{1}\IEEEauthorrefmark{2}}\\
   	    \IEEEauthorblockA{\IEEEauthorrefmark{1}Max-Planck-Institute f\"ur Astrophysik, Karl-Schwarzschild Strasse 1, D-85741 Garching, Germany}\\
    \IEEEauthorblockA{\IEEEauthorrefmark{2}CEA Saclay, Département d'Astrophysique (DAp)/DEDIP
Orme des Merisiers, F-91191 Gif-sur-Yvette, France
    \\\ matteo@mpa-garching.mpg.de}
	}

\begin{document}
\IEEEoverridecommandlockouts
\IEEEpubid{\makebox[\columnwidth]{\copyright2018 IEEE \hfill} \hspace{\columnsep}\makebox[\columnwidth]{ }}
\maketitle
\IEEEpubidadjcol
\begin{abstract}
Current state-of-the-art simulations of accretion flows onto black holes require a significant level of numerical sophistication, in order to allow the three-dimensional modeling of relativistic magnetized plasma in a regime of strong gravity. We present here a new version of the GRMHD code \texttt{ECHO} \cite{Del-Zanna:2007} developed in collaboration with the Max Planck Computing and Data Facility (MPCDF) and the Leibniz Rechenzentrum (LRZ), which employs a hybrid multidimensional MPI-OpenMP coupled with the production of MPI-HDF5 output files \citep{Bugli:2018b}. The code's high degree of parallelization has been crucial for the study of some fundamental properties of thick accretion disks around black holes, in particular the excitation of non-axisymmetric modes in presence of both hydrodynamic and magnetohydrodynamic instabilities.
\end{abstract}

\section{Introduction}
Accretion of magnetized hot plasma onto compact objects (such as neutron stars and black holes) is one of the most efficient mechanisms in the Universe in producing high-energy radiation. The astrophysical sources powered up by such a process include active galactic nuclei (AGNs, \cite{Marconi:2004,Reynolds:2014}), X-ray binaries \citep{Narayan:1995,Fender:2004} and gamma-ray bursts \citep{Woosley:1993,Kumar:2015}, just to cite a few. In particular, a system formed by a central black hole surrounded by a thick torus of plasma is expected to result from the mergers of compact binaries, the gravitational collapse of massive stars and in tidal disruption events  \citep{Rezzolla:2010,Aloy:2000,Coughlin:2014}.

The morphology of the central engine resulting from our current understanding of accretion onto compact objects is quite complex. A high-density main accretion disk, which feeds matter to the center and emits a significant fraction of the source high-energy luminosity, is usually surrounded by a low-density, highly magnetized hot corona which is the perfect site for magnetic reconnection events and particle acceleration. Moreover, strong magnetic fields can lead to the formation of relativistic polar jets along the rotation axis of the system, which launch matter and energy away from the accreting black hole into the interstellar medium. This process can be so powerful (as in the case of AGNs) that in many instances reshapes the global structure of the host galaxy.

Numerical simulations are of paramount importance in modeling such systems, since they are essentially complementary to both analytic models and direct observations. Indeed, numerical experiments represent an invaluable tool to investigate the complex dynamics of the many physical processes involved in accretion (e.g. gravitation, viscous and dissipative effects, radiative and thermal transport, hydrodynamics, magnetic instabilities,  etc.), allowing the study of the regions in proximity of a central compact object which can hardly be resolved via current observational techniques. 

One popular choice used in numerical modeling of accretion disks is the use of General Relativistic Magnetohydrodynamics (GRMHD), which describes the behavior of a conducting fluid in presence of strong gravitational fields. 
Unfortunately, it remains an open question whether or not accretion disk simulations converge to physically reasonable and reliable results (and if so, to what extent).  Especially in presence of magnetic fields (which are a fundamental ingredient in current models of astrophysical flows), one can have dynamic processes taking place at physical length-scales much smaller than the global size of a typical disk orbiting around a black hole. This leads to the question of whether the global numerical simulations are missing some crucial physical process occuring at unresolved small-scales, which nevertheless shapes the large-scale dynamics of the accretion disk. As a consequence, higher resolutions are constantly required to test the completeness of the information retrieved from numerical simulations.

In this work we present the last developments of the GRMHD code \texttt{ECHO} \citep{Del-Zanna:2007}, whose goal is to investigate on the dynamics of magnetized thick tori orbiting around black holes. In Section\ref{sec:echo} we describe the code's general features, while in Section \ref{sec:parallel} we explain  the recent improvements in its parallel scheme.  Section \ref{sec:disk} introduces the results from the application of the code to the study of thick accretion disks, after which we present our conclusions in Section \ref{sec:conc}.

\section{The code}\label{sec:echo}

The code \texttt{ECHO}  (\emph{Eulerian Conservative High Order} scheme ) is a Godunov-type, shock-capturing, finite-difference scheme that integrates the set of GRMHD equations in a conservative form, which guarantees accuracy and stability during the numerical integration (see \cite{Del-Zanna:2007} for a detailed description of the original scheme). 

Along with the equations describing the conservation of rest mass, momentum, and energy densities that characterize the plasma, the system evolves the electric and magnetic fields by integrating Maxwell equations. The diverge-free condition for the magnetic field (which holds analytically, but  not numerically) is enforced by using an upwind Constrained-Transport (CT) scheme \cite{Londrillo:2004}, which ensures the fulfillment of this constraint to machine precision by defining staggered components of the magnetic field at the cell's interfaces rather than the cell's center.

The stationary gravitational background is set by the central black hole, which can either have a spin or not. The code makes use of spherical Kerr-Schild coordinates, which can penetrate the black hole event horizon and reach a region casually disconnected from the rest of the grid. This choice prevents the propagation through the numerical domain of errors introduced at the radial inner boundary. The numerical grid is stretched in both the radial and polar directions to enhance the resolution close to the central black hole and in proximity of the equatorial plane, respectively.

The time integration is performed with a third order standard Runge-Kutta scheme in the ideal GRMHD approximation, while Implicit-Explicit (IMEX) schemes \citep{Pareschi:2005} are used when the non-ideal Ohm's law closure (which includes the effects of turbulent resistivity and/or mean-field dynamo, see \cite{Bucciantini:2013}) is selected. The difference in integration scheme is due to the fact that the evolution equation for the electric field (i.e. Ampere's law) can contain stiff source terms for a sufficiently low value of the plasma's resistivity, requiring therefore an implicit algorithm that can provide numerical stability to the integration. The draw back of the inclusion of resistivity with respect to the ideal GRMHD approximation is of course the need for the time-integration of three additional equations (one per component of the electric field), which sets a more stringent constraint on the affordability of the simulations. The computation of the up-wind fluxes is accomplished with a third-order Parabolic Piecewise Method (PPM), although other schemes (such as TVD, WENO and MPE) are also available in the code's implementation. Hence the code displays an overall third order of accuracy.

In the regions of the numerical domain where the plasma is rarefied and highly magnetized, numerical errors can lead to negative values for the gas thermal pressure. To avoid inconsistencies, then,  the code evolves the conservation law for the specific entropy (which is not strictly conserved in an astrophysical plasma, as it can increase in correspondence of shocks) and use its value to retrieve a positive pressure which is a good approximation of the expected value. This scenario usually occurs in a few points in the disk's atmosphere, which in most cases are dynamically completely irrelevant for the general outcome of the simulation.

\section{Parallelization scheme}\label{sec:parallel}

The latest development of the numerical algorithms in  \texttt{ECHO} concerns a significant upgrade of its parallelization scheme. The main motivation behind this improvement are the computational costs of the three-dimensional simulations required to investigate the development of non-axisymmetric modes in magnetized thick tori around black holes. The original version of  \texttt{ECHO} included only the possibility to parallelize along one axis of the numerical domain (along $x$ for Cartesian grids, along the radius $r$ for cylindrical and spherical ones), since most of the simulations performed with the code were restricted to two-dimensional domains. This choice allows for a maximum number of MPI tasks which is limited by the grid resolution along the direction to be parallelized, and proved to be unsuitable for our study because of the following reason. 

The minimum resolution required by properly resolved magnetized models is  $\sim256^3$ grid points. A one-dimensional MPI decomposition, as in the original version of  \texttt{ECHO}, would allow for a maximum number of MPI tasks given by
\be\label{eq:nmpi}
N^{\mathrm{MPI}}_{\mathrm{max}}=\frac{N_x}{n_{\mathrm{g}}},
\ee
where $N_x$ is the number of grid points along $x$ (including ghost zones), and $n_{\mathrm{g}}$ is the number of ghost cells required by the interpolation and reconstruction routines. In practice, \eqref{eq:nmpi} gives an overestimate, since a significant amount of time is spent by the code communicating the boundaries between MPI tasks, making the effective maximum number of MPI tasks quite small. In the case of a $256^3$ grid, $N^{\mathrm{MPI}}_{\mathrm{max}}\sim20$, which proves to be inadequate to perform a series of sufficiently long 3D GRMHD simulations, especially when a resistive plasma is considered.

\subsection{Multidimensional MPI domain-decomposition}
To overcome the limitation of  \texttt{ECHO}'s original parallelization scheme, we extended it with a multidimensional MPI domain-decomposition. This allows us to increase the maximum number of MPI tasks that can efficiently be used for a given grid resolution and to reduce the time spent during the communication between neighboring cells. Finally, this provides a mean to greatly reduce the total runtime of a production run, making it possible to perform a significant number of tests and actual simulations. 

The starting point of the parallelization scheme is the definition of the topology of a Cartesian grid of MPI tasks, whose dimensionality can range from 1 to 3. Via standard MPI Fortran routines each processor is given a specific rank and coordinates in such a grid, along with a portion of the numerical domain (or \emph{sub-domain}) to work on. 

As long as the code does local calculations (e.g. Runge-Kutta integration, calculation of source terms, etc.) each processor does not need to interact with the others. When the variables need to be reconstructed at the cell interfaces or the staggered magnetic field components need to be interpolated, the calculations require appropriate boundary informations from the ghost cells. The ghost zones that lay at the borders of the numerical domain represent the real \emph{physical boundaries}, and they are filled with different values depending on the particular physical condition required.

For example, let us consider a generic quantity $Q=Q(x_i)\equiv Q_i$ and the problem of selecting the correct values at the inner boundary along the $x$ axis ($i=1$). The ghost zones will correspond then to the grid points $x_{1-i}$ (where $i=1,\dots,n_g$), and we may apply one of the following different prescriptions:
\begin{itemize}
 \item \underline{Positive reflecting boundary}:
 \be
 Q_{1-i}=Q_{i}.
 \ee
 Example: the velocity component parallel to a reflective wall.
 \item \underline{Negative reflecting boundary}: 
 \be
 Q_{1-i}=-Q_{i}.
 \ee
 Example: the velocity component perpendicular to a reflective wall.
 \item \underline{Fixed boundary}: 
 \be
 Q_{1-i}=\tilde{Q}.
 \ee
 Example: the pressure at the surface of a star (which would have $\tilde{Q}=0$).
 \item \underline{Open boundary} (extrapolation of $0^\tx{th}$, $1^\tx{st}$ and $2^\tx{nd}$ order): 
 \begin{align}
 Q_{1-i}=\left\lbrace
 \begin{array}{r@{}l}
 Q_{1}\\
 2Q_{1-i}-Q_{2-i}\\
 3(Q_{1-i}-Q_{2-i})+Q_{3-i}
 \end{array}
 \right..
 \end{align}
 Example: every quantity that is allowed to flow outside the boundary.
\end{itemize}

If the ghost zones are within the numerical domain due to the domain decomposition (or in the special case of periodic boundaries), then each MPI task will have to communicate with all its neighbors to send the values needed by the others MPI tasks for their ghost zones, and  in return receive the values to fill its own boundaries. This \emph{communication phase} is performed along all axis of the numerical domain for each MPI task anytime the boundary values are required. This can take a non-negligible amount of computing time (with respect of the total run-time of the application). Although a multidimensional decomposition allows for a larger maximum number of MPI tasks, it is still recommended to check how computationally expensive (in term of CPU time) the communication phase is. 

A reasonable metric to evaluate a priori whether or not a particular MPI decomposition is advantageous can be obtained by computing the ratio between the volume of data that need to be communicated and the volume of data that form the local sub-domain. Let us consider a grid resolution $N^3$, $n_{\mathrm{g}}$ ghost cells per boundary, and an MPI domain decomposition into $\mathcal{N}$ total tasks. Let us call $n_x$, $n_y$ and $n_z$, respectively, the number of MPI tasks along the $x$, $y$, and $z$ direction, which may differ one from each other, and are equal to 1 in case a direction is not split by the decomposition. The total data-volume that each MPI task will need to communicate for a given dimensionality $d$ is
\begin{align}
V_{\mathrm{ghost}}= 2n_{\mathrm{g}}N^2\times\left\lbrace
\begin{array}{r@{}l}
1&\qquad\mathrm{for\ }d=1\\	 
\left(\frac{1}{n_y}+\frac{1}{n_z}\right)&\qquad\mathrm{for\ }d=2\\ 
\left(\frac{1}{n_xn_y}+\frac{1}{n_xn_z}+\frac{1}{n_yn_z}\right)&\qquad\mathrm{for\ }d=3
\end{array}
\right.
\end{align}
where we assumed a 1-dimensional decomposition done along the $z$ axis and the 2-dimensional one along the $y$ and $z$ axis \footnote{This choice is suggested by the way the Fortran language stores arrays in the computer memory.}. If we now divide $V_{\mathrm{ghost}}$ by the data-volume of the local domain $V_{\mathrm{dom}}=N^3/\mathcal{N}$ we obtain 
\begin{align}
\chi\equiv\frac{V_{\mathrm{ghost}}}{V_{\mathrm{dom}}}=\frac{2n_{\mathrm{g}}}{N}\times\left\lbrace
\begin{array}{r@{}l}
n_z&\qquad\mathrm{for\ }d=1\\	 
\left(n_y+n_z\right)&\qquad\mathrm{for\ }d=2\\ 
\left(n_x+n_y+n_z\right)&\qquad\mathrm{for\ }d=3
\end{array}
\right.,
\end{align}
which can be expressed in compact form as
\be
\chi=\frac{2n_{\mathrm{g}}}{N}\sum_{i=1}^d{n_i}.
\ee
The volume of all communicated ghost zones for a given local domain can equal at most the volume of the domain times twice the dimensionality of the decomposition. If the domain size is shrunk even further, there would not be enough grid points to communicate to the nearest neighbors. Therefore, we define the quantity
\be
\tilde{\chi}=\frac{\chi}{2d}=\frac{n_g}{Nd}\sum_{i=1}^d{n_i},
\ee 
which is normalized to 1 for any decomposition dimensionality. For the parallelization setup to be effective, the value of $\tilde{\chi}$ has to be sufficiently small, otherwise the code will start being \emph{communication dominated} instead of \emph{computation dominated}, i.e. a large fraction of the run-time will be used to exchange the boundary conditions between MPI tasks instead of performing actual calculations.

It is worth mentioning that the new scheme allows to define the dimensionality of the decomposition and the order of the axis along which the decomposition is performed: this feature is particularly relevant for those cases where the numerical domain has in one direction many more point than in the others, therefore requiring very different number of MPI tasks along the various axis.

\begin{figure}[!b]\centering
 \includegraphics[width=0.5\textwidth]{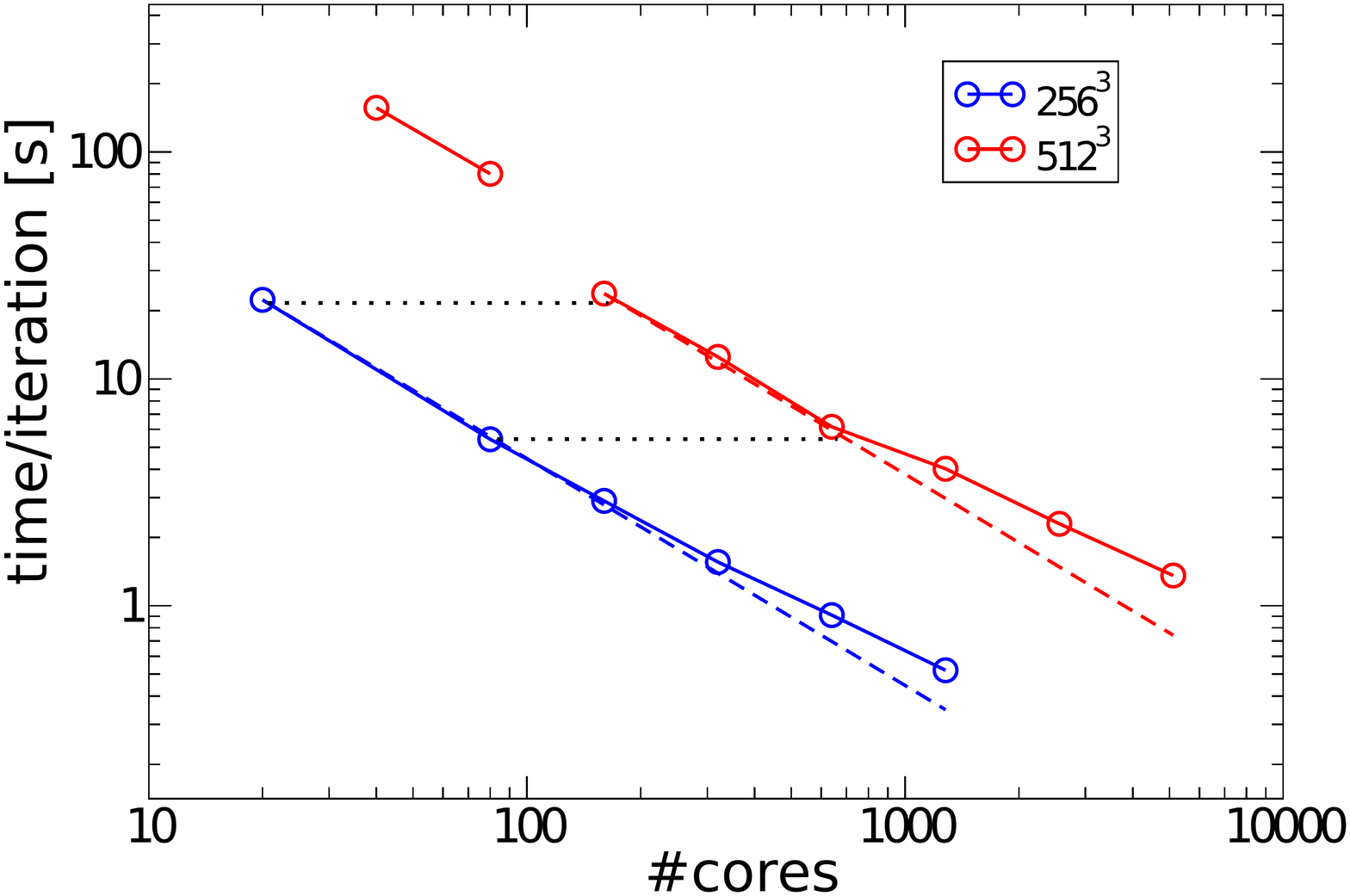}
 \caption[Scaling plot on the Hydra cluster for the Alfv\`en wave test with a 2D MPI domain decomposition.]{Scaling plot on the Hydra cluster for the Alfv\`en wave test with a 2D MPI domain decomposition. The blue curve refers to a $256^3$ grid and the red curve to a $512^3$ grid. The dashed lines represent the ideal \emph{strong} scaling, while the dotted ones correspond to the ideal \emph{weak} scaling.\label{fig:scaling_hydra}}
\end{figure}

\subsection{Scaling}
To quantify the performance of the parallelization scheme, we ran a series of scaling tests with increasing number of MPI tasks and for different grid resolutions.

There are two important characteristics of a well-behaved parallel code that we are interested in:
\begin{itemize}
\item \emph{Strong scaling}: ideally the value of run-time per iteration should be inversely proportional to the number of cores used. Deviations from this trend can be due to a non-optimal implementation of the communication routines. We note that there is always a saturation point beyond which the code becomes communication-dominated, which we mentioned when introducing the $\tilde{\chi}$ parameter.
\item \emph{Weak scaling}: ideally the value of run-time per iteration should stay constant for the same ratio of number of grid zones $N^3$ over number of cores $\mathcal{N}$. This statement represents the expectation that when the volume of workload is increased by a certain factor, increasing accordingly the number of MPI tasks working on it should compensate for it. Deviations from this behavior are usually a good indicator for margins of improvement in the scheme.
\end{itemize}
We estimated the improvement given by a higher dimensionality of the domain decomposition by comparing two different sets of tests, the first one conducted on the Hydra\footnote{The main part of the cluster, which was used for our tests, consists in $\sim3500$ nodes with 20 Intel Ivy Bridge cores @ 2.8 GHz and 64 GB memory each. Visit \url{www.mpcdf.mpg.de/services/computing/hydra/} for more details.} cluster hosted by the \emph{Max Planck Computing and Data Facility} (MPDCF), and the second one on SuperMuc\footnote{The tests were launched on the \emph{Phase1 Thin nodes} of the cluster. Each node mounts 16 Intel Sandy Bridge cores @ 2.7 GHz and 32 GB of memory. More info on \url{https://www.lrz.de/services/compute/supermuc/systemdescription/}.} at the \emph{Leibniz Rechenzentrum} (LRZ). Both kinds of tests involved the solution of a standard MHD problem, i.e. the propagation of an Alfv\`en wave through a three-dimensional domain.

On Hydra we used two different resolutions ($256^3$ and $512^3$ grid points) running on a number of cores that ranged from 20 (corresponding to a single node) up to 5120 (hence 256 nodes) using a 2D MPI domain decomposition. Concerning the numerical algorithms, we opted for a $3^\tx{rd}$ order Runge-Kutta scheme coupled to an MPE5 reconstruction (requiring three ghost cells) and an HLL Riemann solver. The results for the test on the Hydra cluster are shown in \refig{fig:scaling_hydra}. For the lower resolution run the code proved to have a perfect strong scaling up to 160 cores, but started to be communication dominated for larger numbers of cores. The same sort of saturation is exhibited by the high-resolution run, although for a smaller relative number of cores. Overall we obtained a good weak scaling for a ratio of grid points to number of cores down to $N^3/\mathcal{N}=256^3/160\simeq10^5$, then the high-resolution run started to be communication dominated.

\begin{figure}\centering
 \includegraphics[width=0.5\textwidth]{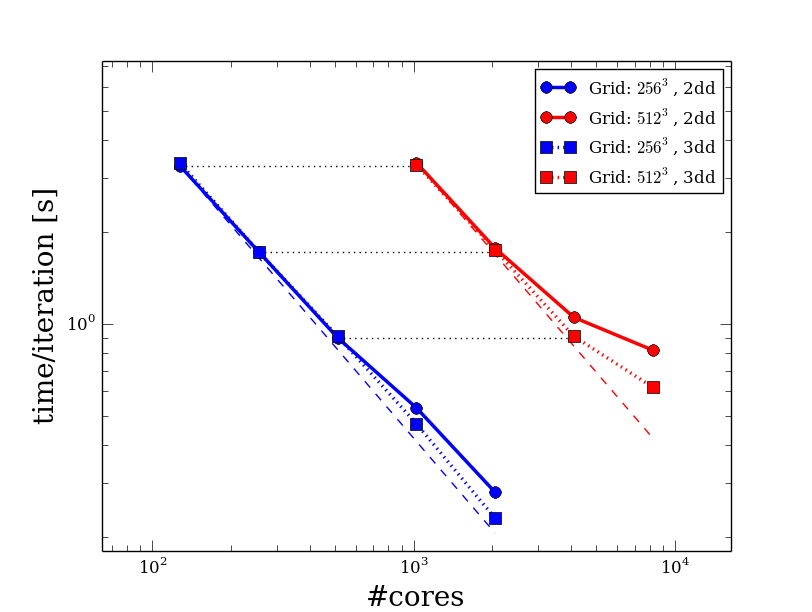}
 \caption{Scalability plot SuperMUC Phase1 for the Alfv\`en wave problem with both a 2D (circles) and 3D (squares) MPI domain decomposition. The blue curves refer to a $256^3$ grid and the red curves to a $512^3$ grid. The dashed lines represent the ideal \emph{strong} scaling, while the dotted ones correspond to the ideal \emph{weak} scaling.\label{fig:scaling_supermuc}}
\end{figure}

On SuperMuc we performed the same physical test but with a few differences. To match the setup of the productions runs involving the evolution of magnetized disks, we used a $2^\tx{nd}$ order Runge-Kutta scheme coupled to a PPM reconstruction (still requiring 3 ghost cells). Furthermore, in addition to the two-dimensional domain decomposition (2DD) tested on Hydra we also checked the performance of the more recent three-dimensional one (3DD). The number of cores ranged from 128 (8 nodes) to 8196 (512 nodes, corresponding to one full island). As we can see from \refig{fig:scaling_supermuc}, the 3DD provides overall faster computation than the 2DD, allowing a nearly perfect weak scaling up to 4096 cores. The runs on the $256^3$ grid show very little saturation up to 2048 cores, beyond which the $\tilde{\chi}$ parameter reaches values greater than $\sim0.27$ for the 2DD and leads to a significant deviation from the ideal strong scaling limit.

\begin{figure}\centering
 \includegraphics[width=0.5\textwidth]{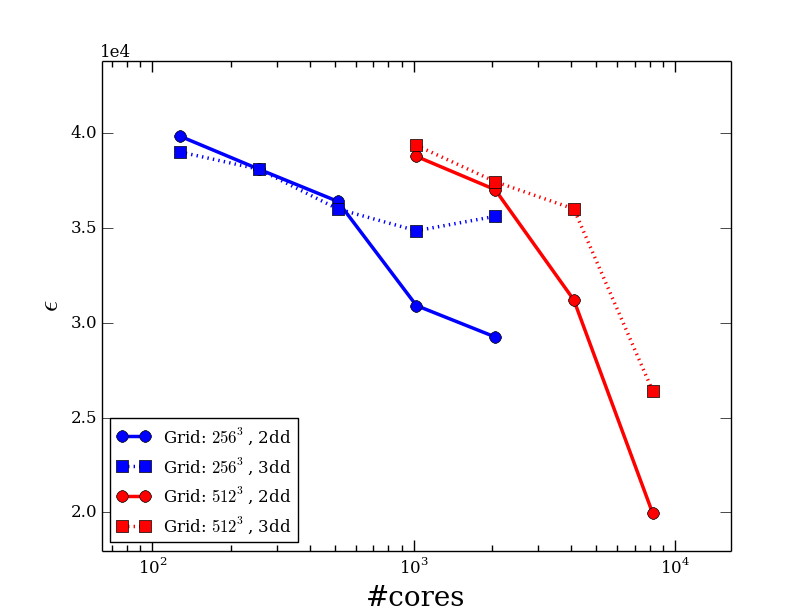}
 \caption[Plot of the code's efficiency $\varepsilon$ vs the number of cores for a 2D and 3D MPI domain decomposition.]{Plot of the code's efficiency $\epsilon$ vs the number of cores for a 2D (circles) and 3D (squares) MPI domain decomposition. The blue curves refer to a $256^3$ grid and the red curves to a $512^3$ grid.\label{fig:efficiency_supermuc}}
\end{figure}
As further metric for the good behavior of the code, we computed for each run the quantity 
\be
\varepsilon=\frac{IN^3}{\mathcal{N}t},
\ee
where $I$ is the number of iterations and $t$ is the runtime of the simulation. Ideally, the ratio between the number of \emph{zone cycles} ($IN^3$) and the \emph{CPU-time} ($\mathcal{N}t$) should be a constant of the code, independent of the particular problem or parallelization scheme. For this reason the parameter $\varepsilon$ can be referred to as the \emph{code efficiency}. As shown in \refig{fig:efficiency_supermuc}, the value of $\varepsilon$ for the 3DD varies much less than in the 2DD case. Apart from the last run at 8196 cores, the 3DD provided a mean value of $\varepsilon\sim3.7\times10^4$ with variations on at most $\delta\varepsilon\sim5\%$. The efficiency of both runs with the 2DD started to decrease significantly and depart from the 3DD case when $\tilde{\chi}\gtrsim 0.27$, once again suggesting that for larger values of $\tilde{\chi}$ the application starts to be dominated by communication. However, the strong decrease in $\varepsilon$ at 1024 cores involves both runs with 2DD and 3DD, pointing therefore to another contingent source of inefficiency for those specific runs. 

\begin{figure}
\includegraphics[width=0.47\textwidth]{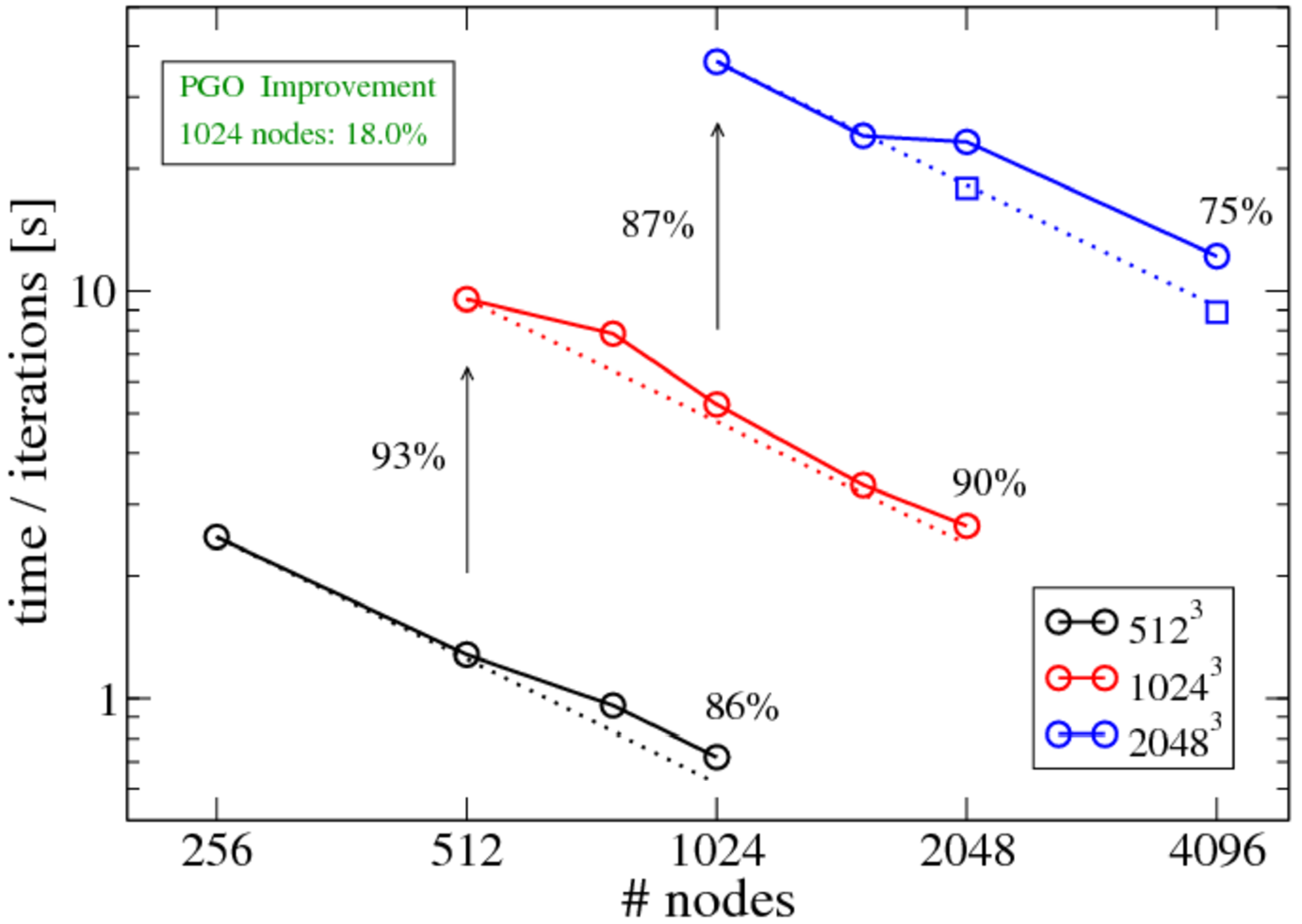}
\caption{Strong and weak scalability plot performed on SuperMUC Phase1, with problem’s size of $512^3$ (black), $1024^3$ (red) and $2048^3$ (blue curve) grid points. The test considered was the propagation of an Alfv\`en wave in a cubic domain. A 3D MPI domain decomposition was used for all tests. The percentages indicate the code’s parallel efficiency in a regime of strong (end point of each curve) and weak (vertical arrows) scaling.\label{fig:scaling}}
\end{figure}

As a last test of the parallelization scheme's performance, we launched a collection of high resolution simulations with $512^3$, $1024^3$ and $2048^3$ grid points on SuperMUC, using the same setup of the previous set of runs but selecting only the 3DD scheme. From \refig{fig:scaling} we can see how the code exhibits very good weak and strong scaling properties from 4096 up to 65536 cores (corresponding to 8 islands on SuperMUC Phase1). A significant speed-up of the code was furthermore achieved by exploiting the Profile Guiding Optimization (PGO) options provided by the Intel compiler, which led to an improve in performance by up to 18\% \footnote{These results were obtained at the \emph{LRZ Scaling Workshop 2017} \cite{Jamitzky:2017}, during which \texttt{ECHO} won the \emph{LRZ Scaling Award 2017} for the best relative improvement in a highly parallelized scientific application.}.

It is important to notice once again that with a one-dimensional MPI domain decomposition the maximum number of cores at disposal would have been of the order of 20 for a typical resolution of $256^3$ grid points. This must be compared with the number of cores allowed by a three-dimensional decomposition, which would be of the order of 2000, determining therefore a speed-up of a factor of 100 with respect to the original version of the code (given the code's good scaling behavior). This clearly shows how important the improvement of the parallelization scheme has been, and in general how a higher dimensionality in the domain decomposition leads to systematic better performances of the code.\\

\subsection{Further improvements}
Another bottleneck that has to be considered when performing 3D MHD simulations is the handling of data output. For a typical non-ideal GRMHD run on a $512^3$ grid a single output file has to contain 11 variables defined on the whole domain (two scalar fields for rest mass density and pressure, three vector fields for fluid velocity, electric and magnetic fields), which for a single precision floating-point accuracy means a file-size of $\sim 5.5 $\ GB. An attempt to write the whole file in a serial fashion using a single \emph{master} MPI task that collects all the data and outputs them into a file  will lead to two major practical problems:
\begin{itemize}
\item The local memory accessible by the processor will in general not be large enough to store all the data, since at the same time the code has a large number of quantities that need to be kept in memory for the next computation steps.
\item Writing on file such an amount of data with a single processor may considerably affect the performance of the code, requiring first a communication phase between all the MPI tasks to collect the data and then the actual writing phase. 
\end{itemize}

For these reasons we decided to fully take advantage of the parallelization scheme used during the code's computation and let each MPI task individually write out its data on the same file. To achieve this goal we adopted the \texttt{HDF5}\footnote{\url{https://www.hdfgroup.org}} standard to write our output files (instead of the original plain unformatted binary files) and made use of the \texttt{HDF5-MPI} library. 
Such an approach allows each MPI task to write its own data on a particular slab of the output file in parallel, significantly decreasing the amount of time that the code spends dealing with the output phase of the simulation.

More recently, an additional layer of parallelization is being added to \texttt{ECHO} by including OpenMP parallel regions within specific routines, effectively leading to a hybrid MPI-OpenMP scheme that should further improve the code's performance. The reason for this choice resides in the fact that a significant (if not dominant) fraction of the time spent into performing computations by a GRMHD code is used in the inversion from \emph{conservative variables} (so called because the set of conservative laws that form the GRMHD equations are expressed in their terms) and the \emph{primitive} variables (which are the fundamental quantities that define the physical state of the plasma, such as the rest mass density, velocity, pressure, etc.). By distributing even more the work-load to multiple OpenMP threads, the code can therefore obtain a significant speed-up. The results of this last stage of the code's improvement will be presented in forthcoming works.

\section{Magnetized accretion disks}\label{sec:disk}

We have recently used the improved version of \texttt{ECHO} (without the OpenMP layer) to investigate in detail the growth of non-axisymmetric modes in three-dimensional magnetized tori threaded by a purely toroidal magnetic field \cite{Bugli:2017}, using the ideal GRMHD approximation. The new performance of the code has permitted the study of a large number of models and hence an investigation on the impact of several different parameters, such as strength of the magnetic field, grid resolution and initial perturbations. 

\begin{figure}
\includegraphics[width=0.49\textwidth]{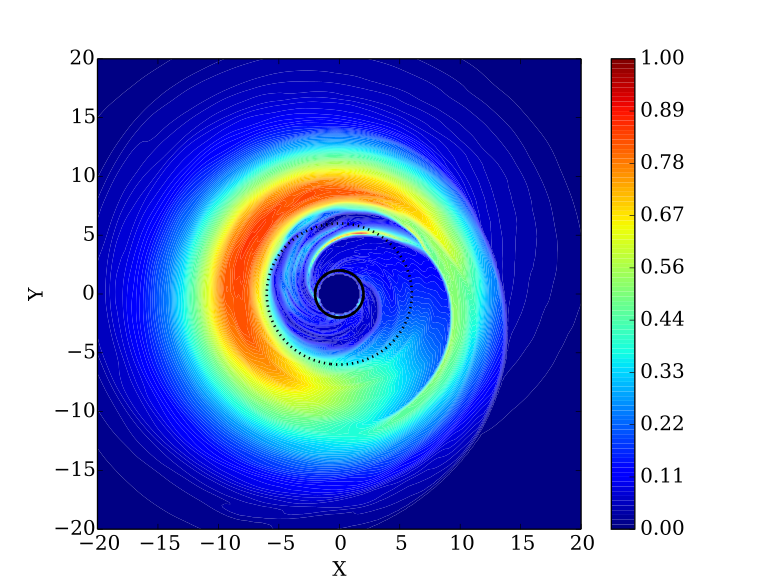}
\includegraphics[width=0.49\textwidth]{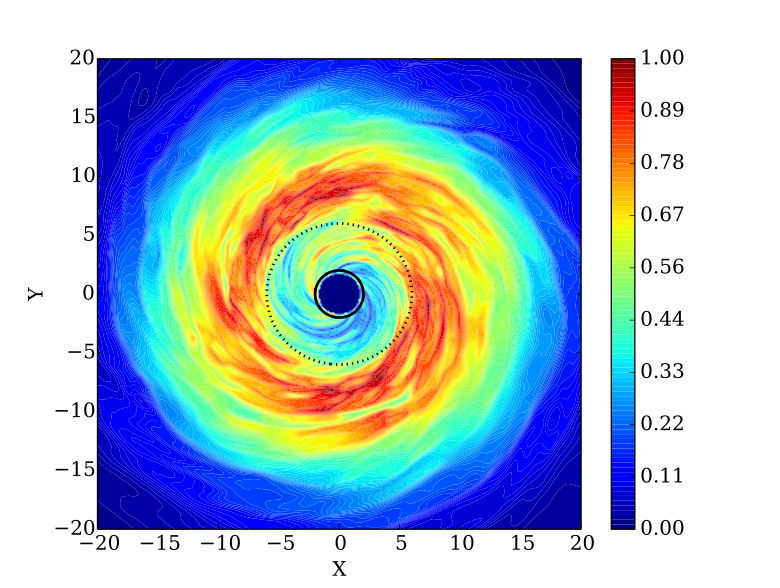}
\caption{Equatorial cut of the rest mass density $\rho$ for model \modela~(top) and model \modelB~(bottom) at the time $t\simeq15\ P_c$. The maximum value of $\rho$ is normalized to 1 in each plot. The solid black curve represents the black hole event horizon, while the dotted curve indicates the radius of the last marginally stable orbit. \label{fig:slices}}
\end{figure}

Selecting as initial condition the analytic magnetized equilibrium provided in \cite{Komissarov:2006}, we studied in detail the interaction of the so-called Papaloizou-Pringle instability (PPI) and the magnetorotational instability (MRI) during both their linear and non-linear stages of development. The PPI is a purely hydrodynamic global instability that affects thick tori and leads to the formation of large-scale azimuthal structures orbiting around the central black hole. For wide disks the PPI fastest growing mode is expected to be the one with azimuthal number $m=1$ (see top panel of \refig{fig:slices}). The MRI, instead, is a local linear MHD instability that commonly appears in most magnetized models of accretion disks (i.e. not only in tori) and develops on dynamical time-scales MHD turbulence and stresses that transport the angular momentum of the disk outwards and therefore leads to significant accretion of matter onto the black hole. In particular, the fastest growing mode of the MRI has in general azimuthal number $m$ much larger than 1, and leads to the growth of fluctuations on scales much smaller than the global size of the disk.

\begin{table}
\caption{Parameters of the models presented in this work.}\label{tab:models}
\centering
\begin{tabular}{cccccc}
\toprule
& $N_r$ & $N_\theta$ & $N_\phi$ & Excitation & $\sigma_c$ \\ 
\midrule
\modela    & 256 & 256 & 256 & Random & 0\\
\modelB    & 256 & 256 & 512 & Random & $10^{-2}$\\
\modelC    & 256 & 256 & 512 & Random & $3\times 10^{-2}$\\
\modelD    & 256 & 256 & 512 & Random & $10^{-1}$\\
\modelA    & 256 & 256 & 512 & $m=1$  & 0 \\
\modelBm   & 256 & 256 & 512 & $m=1$  & $10^{-2}$\\
\midrule
\end{tabular}
\end{table}

Despite the large number of global GRMHD simulations of thick accretion disks conducted in the last two decades by various authors, no evidence of significant development of the PPI  has been found in magnetized models \cite{Hawley:2000,McKinney:2009,Tchekhovskoy:2011,Fragile:2017}. This suggests that the growth of the MRI, by triggering MHD turbulence and transporting angular momentum outwards in the disk, leads the system to a PPI stable configuration and hence prevents the formation of large-scale structures. To what extent this conclusion applies is however not clear, as the vast majority of studies considered initial configurations prone to favor the growth of the MRI fastest growing mode, i.e. using vertical magnetic fields and random initial perturbations. 

To better understand the quenching of the hydrodynamic instability, we consider models threaded by purely toroidal magnetic fields (which trigger the growth of non-axisymmetric MRI, a slower channel with respect to the one selected by vertical fields) and perturb them with both random and monochromatic fluctuations. This choice sets a more favorable environment for the PPI to develop in, and hence allows to investigate whether a scenario dominated by the PPI could still be possible, given appropriate conditions. In \reftab{tab:models} are reported some parameters for the models presented in this work: $N_r$, $N_\theta$ and $N_\phi$ are the number of grid points in radial, polar and azimuthal direction, while $\sigma_c$ is the initial value of the magnetization (ratio between magnetic and thermal pressure) at the center of the disc. We used a $2^\tx{nd}$ order Runge-Kutta scheme for the integration in time, the Harten–Lax–van Leer Riemann solver to compute the up-wind fluxes and the PPM scheme for the reconstruction at interfaces between cells.

\subsection{Results}
\begin{figure}
\includegraphics[width=0.49\textwidth]{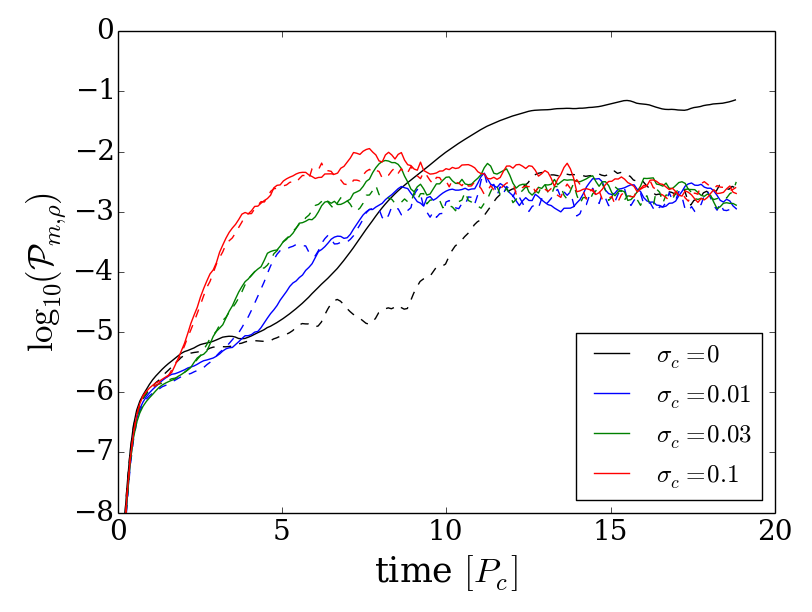}
    \caption{Time evolution of the power in density for the $m=1$ (solid curves) and $m=2$ (dashed curves) modes for models \modela~(black), \modelB~(blue), \modelC~(green), and \modelD~(red).All models are initialized with a random perturbation.\label{fig:power_random}}
\end{figure}

\begin{figure}
\includegraphics[width=0.49\textwidth]{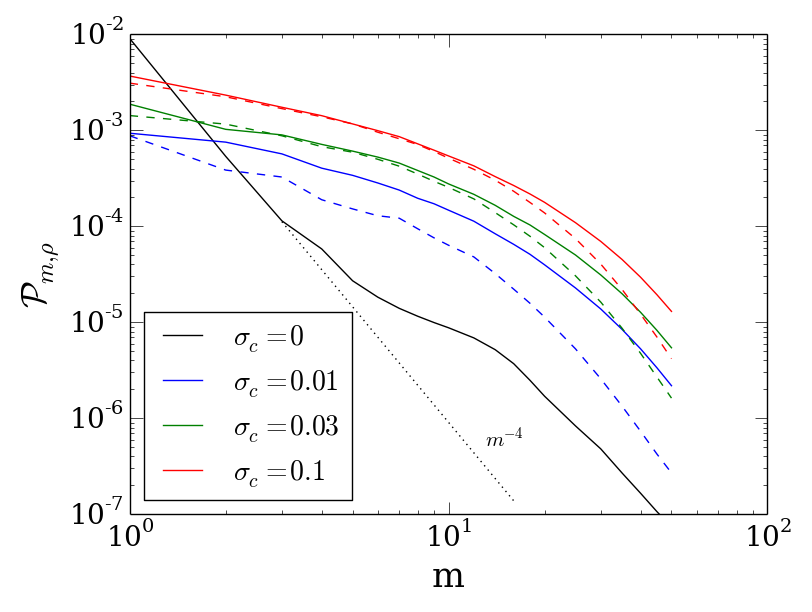}
    \caption{Rest mass density power-spectra in azimuthal number $m$. The black curve represents the hydrodynamic model, the other solid curves refer to magnetized models with high resolution along the $\phi$ direction (512 points) and increasingly high central magnetization $\sigma_c$. The dashed curves stand for low-resolution models (256 points along $\phi$) with the same corresponding value of $\sigma_c$.\label{fig:spectra}}
\end{figure}

We verified that when the magnetized torus is perturbed with a small random fluctuation in the velocity field, the PPI fails to develop significantly, as the concurrent growth of the MRI and small-scale MHD turbulence proceeds faster and prevents the formation of the large-scale modes. From the bottom panel of \refig{fig:slices} it is evident how the action of the MRI halts the formation of a dominant $m=1$ and produces significant accretion of matter onto the black hole. A more quantitative measurement is provided by the power content of the $m=1$ and $m=2$ modes displayed in \refig{fig:power_random}. Only model \modela~exhibits a clear dominance of the $m=1$ mode early in the simulation, which lasts through out the whole saturated phase. For the randomly perturbed magnetized  models, instead, there is no significant difference in power between the two modes, and although a stronger magnetic fields leads to an earlier growth of azimuthal modes, the power saturates to the same level independently of the initial magnetization. These results are consistent with the general missing of large-scale structures in global simulations of magnetized thick disks, and are corroborated by the azimuthal spectra shown in \refig{fig:spectra}. While the hydrodynamic model presents an evident excess of power in the low order modes, peaking at $m=1$ and decaying with a rather steep power law, all the spectra of randomly perturbed magnetized models have much flatter profiles and exhibit much more excited high-order modes.

\begin{figure}
\includegraphics[width=0.49\textwidth]{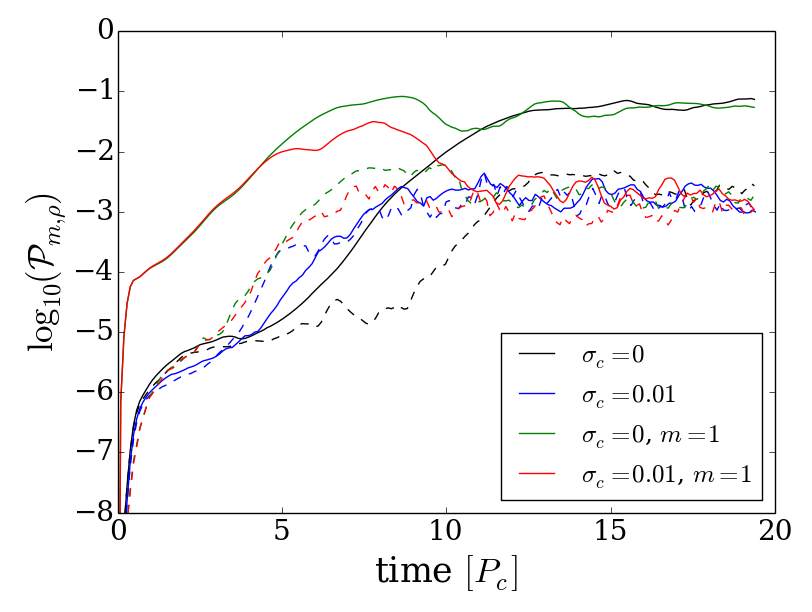}
    \caption{Time evolution of the power in density for the $m=1$ (solid curves) and $m=2$ (dashed curves) modes for models \modela~(black), \modelB~(blue), \modelA~(green), and \modelBm~(red). The former two models are initialized with a random perturbation, while the others are initialized with an $m=1$ perturbation.\label{fig:power_m1}}
\end{figure}

However, when considering a disk with an initially excited $m=1$ mode (model \modelBm) the PPI goes through a transient growth phase similar to the hydrodynamic case, where the PPI fastest growing mode dominates for a significant number of orbital periods $P_c$ of the disk (see the green and red curve in \refig{fig:power_m1}). After this phase, the $m=1$ mode starts being damped by the coupling with the higher order modes excited by the MRI, and its power decreases until it reaches the same value of the randomly perturbed models. The action of MRI, therefore, does not simply halt the growth of the PPI fastest growing mode, but it effectively damps it towards a final turbulent state that is quite independent of the initial perturbation or strength of the magnetic field.

When we considered the evolution of the randomly perturbed magnetized models with a decreased resolution along the azimuthal direction (dashed curves in \refig{fig:spectra}), some excess of power in the $m=1$ mode showed in the setup with the lowest magnetization. By reducing the resolution, in fact, we effectively increased the \emph{numerical resistivity} of the model, which in return quenched the action of the MRI and allowed for a non-negligible growth of the PPI most unstable mode (see the blue dashed line in \refig{fig:spectra}). This result suggests that the inclusion of a physical turbulent resistivity could in principle deeply affect the dynamical evolution of non-axisymmetric modes in thick tori. 

\section{Conclusions}\label{sec:conc}
We presented the most recent numerical development of the GRMHD code \texttt{ECHO}, whose parallelization scheme I/O production have been vastly improved. The new multidimensional MPI domain-decomposition permits to cut the running times by more than two orders of magnitude (compared to the original version of the code), and has led to the implementation of a parallel MPI-HDF5 management of the code's I/O. A further layer of parallelization based on OpenMP is currently being implemented and tested, with an expected further speed-up  of at least a factor of few.

The new implementation has being employed in the study of thick accretion tori threaded by toroidal magnetic fields. We verified that the hydrodynamic instability known as PPI does not ultimately develop significantly in the system, despite the choice of magnetic field topology (the non-axisymmetric MRI modes have slower growth rates than the axisymmetric counterpart) and initial perturbations (the PPI fastest growing mode excited instead of random fluctuations). However, a transient growth of the PPI occurred for several orbital periods when the disk was initialized with a $m=1$ perturbation. The large-scale structure forming at this stage could, in principle, leave a clear signature in the gravitational waves that should be emitted by the over-density of the $m=1$ mode orbiting around the central black hole \cite{Kiuchi:2011,Mewes:2016}.

When a higher degree of numerical dissipation is introduced in the models, the hydrodynamic instability appears to be reinvigorated, as the decrease in strength of the MHD turbulence weakens the coupling that drains power from the large-scales towards the small-scales. Hence, in the future we plan to perform a series of simulations to investigate the role of an explicit physical resistivity by using the generalized Ohm's law closure for resistive plasmas that \texttt{ECHO} can employ.

\section{Acknowledgments}
The author gratefully acknowledges the Leibniz Supercomputing Centre (LRZ; \url{http://www.lrz.de}) and the Max Planck Computing and Data Facility (MPCDF; \url{http://www.mpcdf.mpg.de}) for funding this project by providing computing time, respectively, on the SuperMUC (project PR62LU) and Hydra clusters.

A special thanks goes to F. Baruffa and M. Rampp for their invaluable assistance and support during the development of the parallelization scheme used in ECHO.

\bibliographystyle{IEEEtran}
\bibliography{bugli17}

\end{document}